\documentclass[twocolumn,showpacs,prb]{revtex4}

\usepackage{times,xspace}
\usepackage{amsbsy,amssymb,amsmath,bm}
\usepackage{graphicx,graphics,color,epsfig,rotate}
\usepackage{fancyhdr}

\def\bbbc{{\mathchoice {\setbox0=\hbox{$\displaystyle\rm C$}\hbox{\hbox
to0pt{\kern0.4\wd0\vrule height0.9\ht0\hss}\box0}}
{\setbox0=\hbox{$\textstyle\rm C$}\hbox{\hbox
to0pt{\kern0.4\wd0\vrule height0.9\ht0\hss}\box0}}
{\setbox0=\hbox{$\scriptstyle\rm C$}\hbox{\hbox
to0pt{\kern0.4\wd0\vrule height0.9\ht0\hss}\box0}}
{\setbox0=\hbox{$\scriptscriptstyle\rm C$}\hbox{\hbox
to0pt{\kern0.4\wd0\vrule height0.9\ht0\hss}\box0}}}}

\newcommand{\ignore}[1]{}
\newcommand{\mComment}[1]{}
\newcommand{\gComment}[1]{}
\newcommand{\jComment}[1]{}
\newcommand{\rComment}[1]{}
\newcommand{\lComment}[1]{}

\newcommand{\HeThree}{$^3$He}
\newcommand{\HeFour}{$^4$He}

\renewcommand{\mComment}[1]{\textcolor{red}{Bruce: #1}}
\renewcommand{\gComment}[1]{\textcolor{green}{Zohar: #1}}
\renewcommand{\jComment}[1]{\textcolor{green}{Cristian: #1}}

\begin{document}

\title{On the origin of the decrease in the torsional oscillator period of solid \HeFour}

\author{Z. Nussinov $^{1}$, A. V. Balatsky$^{2}$, M. J. Graf$^{2}$, and S. A. Trugman$^{2}$}

\affiliation{$^1$ Department of Physics, Washington University, St. Louis, MO 63160 USA}
\affiliation{$^2$Theoretical Division, Los Alamos National Laboratory, Los Alamos, New Mexico 87545, USA}
\date{June 2007, to appear in Phys. Rev. B}

\begin{abstract}
A decrease in the rotational period observed in torsional oscillator
measurements was recently taken as a possible indication of a
putative supersolid state of helium. We
reexamine this interpretation and note that the decrease in the
rotation period is also consistent with a solidification of a small
liquidlike component into a low-temperature glass. Such a
solidification may occur by a low-temperature quench of topological
defects (e.g., grain boundaries or dislocations) which we examined
in an earlier work. The low-temperature glass can account for not
only a monotonic decrease in the rotation period as the temperature
is lowered but also explains the peak in the dissipation occurring
near the transition point. Unlike the non-classical rotational
inertia scenario, which depends on the supersolid fraction, the
dependence of the rotational period on external parameters, e.g.,
the oscillator velocity, provides an alternate interpretation of the
oscillator experiments. 
\end{abstract}

\pacs{73.21.-b}

\maketitle
\section{Introduction}

Supersolidity is a unique state of matter that simultaneously displays
both superfluidity and crystalline order.
Much of our understanding of this exotic state is based on the
pioneering work of Penrose and Onsager,\cite{Penrose1956}
Andreev and Lifshitz,\cite{AL}
Chester and Reatto,\cite{Chester67,Reatto69,Chester70}
Leggett,\cite{Leggett70} and Anderson.\cite{Anderson84}
Recently, Anderson and coworkers revisited this problem.
\cite{Anderson05,ABH} Current developments in this field,
presented at a KITP workshop at Santa Barbara,
are available online.\cite{KITP}

\HeFour\ has long been thought
to be the most likely candidate for the supersolid state. Torsional
oscillator experiments by Kim and Chan \cite{Chan04,Kim05,Kim05b,Chan:05}
generated renewed interest in this possibility. \cite{Beam,Legget}
In addition to the work of Kim and Chan, \cite{Chan04,Kim05,Kim05b,Chan:05}
there is now an independent confirmation of the anomalous behavior
of solid \HeFour, as presented by the groups of Reppy and
Shirahama.\cite{rep,Sophie06,Shirahama06,Kondo06,Sophie07}
All of these groups use torsional oscillators similar to those employed by
Kim and Chan.\cite{Chan04,Kim05,Kim05b,Chan:05} Rittner and Reppy
\cite{Sophie06} report history dependence of the signal, when
annealing the sample, to the extent of no observation of any mass
decoupling in the torsional oscillator experiment.
On the other hand, rapid freezing of helium leads
to disorder and a drastic increase of the signal.\cite{Sophie07}
Taken at face value, these torsional
oscillator experiments indicate an anomalous mechanical
behavior of solid \HeFour\ at low
temperatures. Nevertheless, the connection between the mechanical
measurements and the suggested  supersolidity remains tenuous.
The case for supersolidity entails a certain assumption, which
we will discuss at length below, for the interpretation of the
oscillator
data.

The measurements by 
Refs.~\onlinecite{Chan04,Kim05,Kim05b,Chan:05,Sophie06,Shirahama06,Kondo06,Sophie07}
monitor the period of torsional oscillators in the presence
of an applied torque. The purpose of our work is to examine
the mechanical properties of the torsional oscillator.
We will only allude to linear response theory and
causality (the Kramers-Kronig relations).
We show that
the change in torsional oscillator period
may be triggered by a change
in damping and other oscillator parameters.
Such a change may
result from a low-temperature quench of residual
topological defects in the solidification into
a glass. 
Here, we will use the term ``liquidlike'' component to describe elastic defects
that are not pinned above a glass transition at high temperatures. 
In Ref.~\onlinecite{us}, we first explained how a low-temperature pinning of 
elastic defects (two-level tunneling systems) can account
for the reported excess low-temperature specific heat with regards to a perfect Debye crystal. 
The results of Ref.~\onlinecite{us} are at odds with a {\it simple} uniform
supersolid transition, but can also occur in an exotic
glassy supersolid state as discussed by Boninsegni et al. \cite{superglass}
or Wu and Phillips\cite{Wu07}. The effects
described in this paper may account for
the observed change of the torsional oscillator
period. The central thesis
of our current work is that there is no imperative to adduce 
a supersolid fraction which depends very
sensitively on a host of external parameters 
in order to explain the reduction in the 
torsional oscillator period as temperature
is lowered. Rather, the observed
trends might be a very natural
and quite universal outcome of a 
glass transition (regardless of 
its classical or quantum 
nature).

A direct proof of superfluidity would be the observation of persistent
current. In this regard, we note a recent experimental search for
superflow by Day et.~al.,\cite{Beamish05,Beamish06}
who found no mass
flow of any kind to very high accuracy. Relying on these results
and, most notably, on a new thermodynamic analysis, we
recently concluded \cite{us}
that the effect first observed
\cite{Chan04,Kim05,Kim05b} is likely not an intrinsic property of solid \HeFour. This is so because
the low-temperature behavior depends critically on
the \HeThree\ concentration, shows annealing and rapid freezing
dependence, \cite{Sophie06,Sophie07} and,  most importantly,
has a specific heat which is inconsistent with a
simple supersolid transition of a 1\% condensate
fraction.\cite{us} However, our earlier work \cite{us} does not rule
out a supersolid fraction of order ${\cal{O}}(10^{-5})$.

\section{Outline}

Our principal findings are briefly summarized
in Sec.~(\ref{import}). To set the stage for these results,
we will briefly review, in Sec.~(\ref{NC}), the non-classical
rotation inertia (NCRI) effect. We then discuss
the current possible ground-breaking implication
to the observations of
Refs.~\onlinecite{Chan04,Kim05,Kim05b,Chan:05,Sophie06,Shirahama06,Kondo06,Sophie07}
[wherein the change in torsional oscillator period may be
triggered by a supersolid phase]. We then step
back in Sec.~(\ref{how})
and analyze the results for a first principle
linear response analysis without assuming the NCRI. Consequently,
we show how the experimental results 
are consistent with a change in damping
or other oscillator parameters
which are brought
about by the solidification of a small liquidlike
component
as the temperature is lowered. In Sec.~(\ref{sssolid}), we examine
the experimental consequences of a uniform supersolid
transition and illustrate that a simple
supersolid transition cannot
account for the peak in the dissipation
observed by Ref.~\onlinecite{Sophie06}. We then determine,
in Sec.~(\ref{glass_section}),
the angular response for a glass transition,
which can account for all of the existing
trends in the data, derive in detail
our new results on the dissipation and torsional oscillator
period for a glass, and suggest an experiment to distinguish
between the glass versus supersolid scenario.
We conclude in Sec.~(\ref{conc}) with a brief summary
of our results and their relevance to the
torsional oscillator
data of Refs.~\onlinecite{Chan04,Kim05,Kim05b,Chan:05,Sophie06,Shirahama06,Kondo06,Sophie07}.
In the appendix (Sec.~\ref{alts}),
we illustrate that other, non-glass,
response functions can also account for
the data without
assuming a supersolid transition.

\section{Principal Results}
\label{import}

We will address the linear response theory of a driven
torsional oscillator and derive our key results in  Sec.~(\ref{glass_section}).
A general treatment of the problem involves the analysis of possible
modes in the vessel which are excited by the external
torque and their influence on the torsional oscillator.
Depending on their character, such modes can easily
be disrupted by the insertion of barriers and
lead to a far smaller difference between the response
of a liquidlike component and that of a glasslike component.
To make the discussion clear, we will focus on a
simple form for these modes,
which are characteristic
of glass formers. Here we will show:

A glass transition, irrespective of
whether or not it appears in a normal solid
or supersolid, leads to specific
testable predictions for the oscillator period
and dissipation.\cite{superglass,Wu07}
Unlike a {\it simple} supersolid
transition, i.e., a transition where only a change in the inertia of
the oscillator occurs, our key result Eq.~(\ref{gfinalg})
predicts that, as the temperature
is lowered, two principal trends will
be noted: (i) a monotonic decrease
in the oscillator period and
(ii) a peak in the dissipation.
A uniform, mean-field type, supersolid
transition cannot account for phenomenon (ii).

It is important to
stress, that unlike all of the works to
date, we do not assume that only
a supersolid can account
for the observed decrease
in the oscillator period [feature (i)].
Indeed, as we show for the first time,
glassy relaxations mandate the
current observations of features
(i) and (ii).
This occurs universally and is
independent of the specific microscopic
mechanism by which relaxation
processes occur or the detailed
character of the glass. It is nevertheless
encouraging to note that samples in
which the most pronounced anomalous behavior
is observed contain defects.\cite{Sophie07,barometer}

We considered other (non-glass) ``freezing'' transitions that
can also account for the empirical trends. To make this concrete, in
the appendix A, we will detail how a decrease in damping (as a
result of a freezing transition) can also account for the observed
tendencies.
Unlike for the results discussed above,
here we cannot
provide a specific functional form to fit to. To provide the
simplest qualitative form, we invoke Gaussian distributions for the
quantities in question.

Finally, we note that throughout our work, our general form for the
angular response allows for a dependence on
external parameters such as the initial oscillator velocity. These
parameters, e.g. stiffness and density and cell torsional
frequency, which parameterize the external torques, are
convolved
with the intrinsic angular response.
A {\it liquidlike} component in the cell may respond
differently to different initial oscillator velocities.
Such a dependence is observed experimentally. Aside from other
possible effects, external parameters [whether in a supersolid, a
glass, or in any other phase] will also always appear via their
incorporation in the external torques in the general form for the
angular response.

\section{Non-classical Rotational Inertia in a nutshell}
\label{NC}

The basic idea underlying the torsional oscillator measurements
is a test of the non-classical rotational inertia (NCRI) - the
inability of a superfluid to rotate under an applied torsional
drive.
The analog of this experiment on liquid \HeFour\ was performed
by Andronikashvili. \cite{Andronikashvili46,Andronikashvili66}
Later Hess and Fairbank \cite{HF} verified
the existence of the NCRI and of a rotational Meissner effect
for superfluid helium, which was predicted by London. \cite{London}
As a consequence of a superfluid state, which no longer
rotates with the ``normal'' liquid,
the effective moment of inertia
decreases as the superfluid fraction $f_{s}$ increases
\begin{eqnarray}
I_{eff} = I_{osc} + I_{He} [1 - f_{s}(T)]
\label{LondonE}
\end{eqnarray}
which is smaller than the "classical" moment of intertia
$I=I_{osc}+I_{He}$ of the
combined oscillator-helium system.
For an ideal torsional oscillator of stiffness $\alpha$, a measure of
the period ($P = 2 \pi \sqrt{I_{eff}/\alpha}$) indicates what the
superfluid fraction $f_{s}$ is.
The inferred results for $f_{s}$ \cite{HF} agree well with the
superfluid fraction measured by other probes.
The current situation for solid \HeFour\          is far from
being as clear cut. The experiments 
show that
the rotational motion under an applied torque becomes more rapid
at lower temperature- suggesting an analogue of the NCRI for a
{\em supersolid} phase of helium. In this paper
we point out
that the measurements are also consistent with
less exotic material effects, which can
be distinguished by further experiments.

\section{How can we simply understand the
change in torsional oscillator period if it
is not associated with a supersolid?}
\label{how}

The NCRI interpretation of the existing data
is very simple and suggestive. In what follows,
we present a very brief mechanical
analysis of the measurements which points
to an equally simple explanation.
To keep the analysis as general
as possible, we rely on first
principles alone.
The torsional oscillator
experiments 
measure the susceptibility - they do not
directly monitor the moment of inertia of the supersolid.
We start by writing down the general equations of motion
for a torsional oscillator defined by an angular coordinate
$\theta$,
\begin{eqnarray}
[I_{osc} \frac{d^{2}}{dt^{2}} + \gamma_{osc} \frac{d}{dt} + \alpha]
\theta(t) =
\nonumber \\
= \tilde{\tau}_{ext}(t) + \int \tilde{g}(t,t';T) \theta(t') dt'.
\label{de}
\end{eqnarray}
Here, $I_{osc}$ is the moment of inertia of the torsional oscillator,
$\alpha$ is its restoring constant, $\gamma_{osc}$ is the dissipative
coefficient of the oscillator, $\tilde{g}$ arises from
the back action of solid \HeFour\ (it plays the role of a
``polarization''), and $\tau_{ext}$ is the
externally imposed torque. In general,
$\tilde{g}$
is temperature ($T$) dependent.
The oscillator angular coordinate
$\tilde{\theta}(t)$ is a convolution of the applied
external torque $\tilde{\tau}_{ex}$ and the response
function $\tilde{\chi}(t,t')$,
\begin{eqnarray}
\tilde{\theta}(t) = \int dt'
~\tilde{\chi}(t,t') \tilde{\tau}_{ext}(t').
\end{eqnarray}
Causality demands that
$\tilde{\chi}(t<t') =0$. This implies that $\tilde{\chi}(t,t') =
\theta(t-t') \tilde{\chi}(t,t')$; under a Fourier
transformation, this leads to the Kramers-Kronig relations
which we will briefly touch on later. As in any
time translationally invariant system, the Fourier
amplitude of the angular response of the torsion oscillator is
a product
\begin{eqnarray}
 \nonumber
\\  \theta(\omega) = \chi(\omega) \tau_{ext}(\omega),
\label{ft}
\end{eqnarray}
with $\chi = \chi_{1}+ i \chi_{2}$ the angular
susceptibility and $\tau_{ext}$
the external torque in Fourier space. For the simple torsional
oscillator, \begin{eqnarray}
\chi^{-1}(\omega) =
[\alpha - i \gamma_{osc} \omega - I_{osc} \omega^{2}- g(\omega;T)].
\label{central}
\end{eqnarray}
Here, $g(\omega,T)$ is the Fourier transform of the real time
$\tilde{g}$ of Eq.~(\ref{de}).
 For an ideal normal
solid, with a moment of inertia equal to $I_{ns}$, which rotates
with the oscillator, the back action $g(\omega; T)$ is
\begin{eqnarray}
g_{ss}(\omega; T) =   i \gamma_{He} \omega + I_{ns}(T) \omega^{2}
\label{ssberg}
\end{eqnarray}
with $I_{ns}(T)$ the normal component of solid
\HeFour, which varies with temperature
and the dissipation $\gamma_{He}$ is constant.
The moment of inertia contribution of the helium
is orders of magnitude smaller than
that of the empty vessel - the changes
observed in the period are very small
and are a remarkable experimental triumph.
The subscript in $g_{ss}$ refers to the supersolid
interpretation of the results - only a fraction
of solid helium (the ``normal part'')
rotates in unison with the oscillator.
Eq.~(\ref{ssberg}) is consistent with the
NCRI form of Eq.~(\ref{LondonE}). However, we do not
need to impose this form on the existing data.
Current experiments measure the oscillator
period  $P = 2 \pi/ \omega_{0}$  with $\omega_{0}$
the real part of the solution of
\begin{eqnarray}
\chi^{-1}(\omega_0)=0
\end{eqnarray}
at fixed $T$. \cite{explain_period}
For example, a decrease in the dissipative component
$\gamma_{He}$ in the back action $g(\omega; T)$ with
\begin{eqnarray}
g_{diss}(\omega; T) = i \gamma_{He}(T) \omega + I_{ns} \omega^{2} \,,
\label{glassE}
\end{eqnarray}
as the temperature is lowered,
will also lead to a shorter rotation period.

This observation offers a qualitatively
different interpretation of the experimental
results \cite{Chan04,Kim05,Kim05b,Chan:05,Sophie06,Shirahama06,Kondo06,Sophie07}
with no need to invoke supersolidity.
Note that {\it no} superfluid phase is needed to explain the data and there is {\it no} NCRI effect;
the moment of inertia $I_{ns}$ is temperature independent.

The origin of the subscript in
Eq.~(\ref{glassE}) alludes to a
scenario wherein a higher temperature
mobile liquidlike component of the sample
``freezes'' at lower temperatures leading to
a decrease in the dissipation $\gamma$.
In the appendix, we examine in detail
the consequences of Eq.~(\ref{glassE}).
To make our discussion more lucid,
we remark that a qualitatively similar effect appears
in the rotation of hard versus soft boiled
eggs. The solid hard boiled egg
rotates faster than the, liquidlike,
soft boiled egg.

The measured decrease in the rotation period found 
\cite{Chan04,Kim05,Kim05b,Chan:05,Sophie06,Shirahama06,Kondo06,Sophie07}
only implies a crossover in $\chi$ (and a
constraint on $g$). As the real and imaginary parts of $\chi$
are related by the well known Kramers-Kronig (KK) relations,
\begin{eqnarray}
\chi_{1}(\omega) = \frac{2}{\pi} P \int_{0}^{\infty} d\omega' ~  \frac{\omega'
\chi_{2}(\omega')}{\omega'^{2} - \omega^{2}} \nonumber
\\ \chi_{2}(\omega) = - \frac{2 \omega}{\pi}  P \int_{0}^{\infty} d\omega'
~ \frac{\chi_{1}(\omega')}{ \omega'^{2} - \omega^{2}},
\end{eqnarray}
an enhanced decrease in $\chi_{1}(\omega)$
often appears with a pronounced peak
in $\chi_{2}(\omega)$.
A nonvanishing $\chi_{2}(\omega)$ at finite frequency mandates dissipation.
We expect the data to indicate an increase in dissipation concurrent
with the reduction of torsional oscillator period. Whether
or not experimental data adhere to the KK relations is
a powerful check that needs to be done.

In the sections that follow, we explain in some
depth how the supersolid and glass pictures can
both account for the decrease in the rotational
period when the temperature is lowered.
We further examine the dissipation.

\section{A Supersolid origin}
\label{sssolid}

We now determine the experimental consequences
of a simple supersolid transition in which
the moment of inertia follows Eq.~(\ref{LondonE})
(the well known NCRI effect). Here,
the response is given by Eq.~(\ref{ssberg}).
We show  that while a mean-field type NCRI effect
can, as is well known, account for the decrease of the rotation period it
cannot account for the peak in the dissipation $Q^{-1}$ observed by Ref.~\onlinecite{Sophie06}.

We start by reviewing results for an underdamped
harmonic oscillator. The temperature dependent period
of a general damped oscillator
of combined effective moment of inertia $I_{eff} = I_{osc} + I_{ns}$, \cite{i_eff}
stiffness $\alpha$, and dissipation $\gamma$ is
\begin{eqnarray}
\theta(t) = Re\{ A \exp[-i \omega_{0} t - \kappa t]\}
\end{eqnarray}
with $A$ a complex amplitude. For an underdamped oscillator,
the period
\begin{eqnarray}
P \equiv
\frac{2 \pi}{\omega_{0}} = \frac{ 4 \pi I_{eff}(T)}
{\sqrt{4 \alpha I_{eff}(T) - \gamma^{2}(T)}},
\label{periodg}
\end{eqnarray}
\newline
and the
damping rate of the oscillation amplitude
\begin{eqnarray}
\kappa (T) = \frac{\gamma}{2I_{eff}}.
\label{dampingg}
\end{eqnarray}
The ``quality factor'' $Q$, which monitors the number
of oscillations required for a system to have its
energy drop by a factor of $e^{2 \pi}$, is
\begin{eqnarray}
Q = \frac{\sqrt{\alpha I_{eff}}}{\gamma}.
\label{Qfac}
\end{eqnarray}
We can rewrite the changes in oscillator period
as result of small
changes in $\gamma, I_{eff}$ as
\begin{eqnarray}
\delta P = \Big[ \big(
\frac{4 \pi}{\sqrt{4 \alpha I_{eff} - \gamma^{2}}} -
\frac{8 \pi I_{eff} \alpha}{(4 \alpha
I_{eff} - \gamma^{2})^{3/2}}\big) \delta I_{eff} \nonumber
\\ + \frac{4 \pi \gamma I_{eff}}
{(4 \alpha I_{eff} - \gamma^{2})^{3/2}}  \delta \gamma
- \frac{2 \pi I_{eff}^{2}}{(4 \alpha I_{eff} - \gamma^{2})^{3/2}}
\delta \alpha \Big ].
\label{t_expand}
\end{eqnarray}

We see from Eq.~(\ref{t_expand}),
for damping $\gamma < \sqrt{2 I_{eff} \alpha}$,
that the observed decrease in the rotation
period, i.e., $\delta P <0$]
as the temperature
is decreased may be explained by either 
(1) a purported supersolid transition with $g=g_{ss}$, where $\delta I_{eff} <0$ and
$\delta \gamma =0$; 
(2) a (non-glass) freezing transition with a simplified $g=g_{diss}$, where $\gamma_{He}$
becomes smaller as the temperature is decreased, i.e.,
$\delta \gamma <0$ and $\delta I_{eff} =0$, for more details see the appendix~\ref{alts};
(3) a glass transition with an effective increase in the stiffness constant, where
$\delta \alpha>0$ as temperature is lowered.
The glass response to be studied later in Sec.~\ref{glass_section} emulates, in part, the
features of (3).

First, we will consider
a supersolid in which $g=g_{ss}$ of Eq.~(\ref{ssberg}),
$\gamma = \gamma_{osc} + \gamma_{He}$ is fixed and $I_{eff}(T)$
varies with temperature. Later on, we will comment
on the effect of a drop in dissipation expected to
accompany the supersolid state.
For $g=g_{ss}$, we can deduce $I_{eff}(T)$ and then
predict the damping rate $\kappa_{ss}(T)$. In this case,
the damping rate
\begin{eqnarray}
\kappa_{ss} = \frac{\gamma_{osc} + \gamma_{He}}{2(I_{osc} + I_{ns}(T))} \nonumber
\\ = \frac{4 \pi^{2} \gamma}{\alpha P^{2} +
\sqrt{(\alpha P^{2})^{2} - 4 \pi^{2} \gamma^{2} P^{2}}},
\label{modelB}
\end{eqnarray}
or, equivalently,
\begin{eqnarray}
Q_{ss} =
\sqrt{\frac{\alpha^{2} P^{2} + \sqrt{\alpha^{4} P^{4} - 4 \pi^{2}
\alpha^{2} \gamma^{2} P^{2}}}{8 \pi^{2} \gamma^{2}}}.
\label{qb}
\end{eqnarray}
In Eqns.~(\ref{modelB}, \ref{qb}), only
the period $P$ varies
with temperature.
In this simple supersolid picture,
the Q factor is monotonic
in temperature. If $I_{eff}$ decreases as
the temperature is lowered, then the $Q$ factor will
decrease monotonically as the temperature is lowered.

A simple supersolid transition is inconsistent
with the data \cite{Sophie06} where a pronounced
peak in $Q^{-1}$ is seen near
the transition.
A concurrent monotonic drop in the dissipation
is expected to accompany the supersolid transition.
If the drop in the dissipation $\gamma$ is proportional to
the drop in the effective moment of inertia
then $Q^{-1}$ will still be monotonic.
Here, $Q^{-1}$ may increase as $T$ is
lowered. Such an amended
simple supersolid picture is still inconsistent
with the observed peak of $Q^{-1}$. \cite{Sophie06}
Obviously, a nonuniform change in both $\gamma(T)$ and $I_{eff}(T)$
can be engineered to account for the data. Similarly,
a broadening of the transition parameters, similar
to that discussed in the appendix, can lead
to a peak in $Q^{-1}$.

\section{A glass origin}
\label{glass_section}

We next illustrate how a glass transition
can account for the observed data. A
glass transition will not only
give rise to a decrease in the period
but, unlike the supersolid scenario, will
also lead to a peak in the dissipation observed
by Ref.~\onlinecite{Sophie06}.

The external torque is the derivative
of the total angular momentum,
\begin{eqnarray}
L(t) = \frac{d}{dt} \int d^{3}x ~\rho(\vec{x}) r^{2} ~\frac{d}{dt} \theta(\vec{x})
\end{eqnarray}
where $r$ is the distance to the axis of
rotation, $\rho(\vec{x})$
is the mass density and $\frac{d}{dt} \theta(\vec{x})$ the local angular
velocity about the axis of rotation. The experimentally
measured quantity is the angular motion of the torsional oscillator -
not that of the bulk helium,
which is enclosed in it. Ab
initio, we cannot assume that the medium
moves as one rigid body. Similar to a vessel
partially filled with a liquid component, e.g. a soft boiled egg,
an initial imparted external torque can lead to a
differential rotation between the outer torsional
oscillator and a liquid within it. If the
liquid ``freezes" into a glass the medium will move with greater
uniformity and speed. This may lead
to an effect similar to that of the NCRI,
although its origin is completely
different. Furthermore, as a function
of initial rotation speed, the
variance between the response of
``hard" glasslike and ``soft" liquidlike
media changes as temperature is lowered.
Such a difference is indeed observed
in the torsional oscillator experiments
on \HeFour. Our explanations
do not require the supersolid
fraction to depend on the rotation
speed in order to account for the data.

In what follows, we will analyze the
effective equation of motion for
the torsional oscillator. Excited modes
within the medium in its liquid phase
can lead to additional dissipative
torques acting on the torsional oscillator.
These modes can become far slower and
effectively disappear as the system
freezes into a glass state.
Depending on the character of these
modes, the insertion of a barrier
may or may not lead to a change
in the angular motion of the
torsional oscillator. For modes
corresponding to internal flows
around the axis of rotation, such
barriers will disturb the liquidlike modes. Such constrictions will lead to a far
smaller difference between
the period of rotation
of the glass and the
period of rotation with
liquidlike component of \HeFour, if the modes are local.

Henceforth, we will focus on conventional
forms for glasslike relaxations which might be of
use for fitting data. Glasses are characterized by
universal response functions over
a wide range of frequencies and temperatures.
We assume that the additional dissipative modes triggered
by rotation have a similar
form. To be specific, we will assume
that $g(\omega; T)$ obeys the simple dependence as seen
in the far better measured dielectric response
functions (where it reflects the polarization)
of glasses. It is given by a Dyson-like equation, which has been
invoked in Eq.~(\ref{central}),
\begin{eqnarray}
\chi &=& \chi_{0} + \chi_{0} g_{gl} \chi_{0} + ... =
\chi_{0}[1 - g_{gl} \chi_{0}]^{-1} , \nonumber
\\ \chi^{-1} &=& \chi_{0}^{-1} - g_{gl}.
\label{Dyson}
\end{eqnarray}
Here,
\begin{eqnarray}
\chi_{0}^{-1} = [\alpha -
i \gamma \omega -I_{eff} \omega^{2} ]
\label{normalres}
\end{eqnarray}
is the inverse susceptibility of the underdamped system formed by the
torsional oscillator chassis and normal
solid, while $g_{gl}$ is the response function of the overdamped glass component.
Physically, the first term ($\chi_{0}$) in Eq.~(\ref{Dyson})
corresponds to the response of
the bare oscillator to an applied torque, the possible
excitation (by the rotating oscillator) on a transient
mode in the liquid/glass medium, which then acts back on the torsional
oscillator at a later time ($\chi_{0} g_{gl} \chi_{0}$), and all similar
higher
order processes. Employing the standard overdamped form for
glass response functions \cite{Phase1,Phase2,Phase3}
\begin{eqnarray}
g_{gl}=g_{0} [1- i \omega s]^{-\beta},
\label{CD}
\end{eqnarray}
we will be able to provide an
explicit form for $\chi^{-1}$ and thus for all
periods and dissipation as a function of temperature.

In Eq.~(\ref{CD}), $s$ is the characteristic equilibration time
of the liquidlike component when it is perturbed. This time $s$ diverges
in the glass phase (at temperatures $T<T_{0}$)
and is nearly vanishing at very high temperatures.
Similarly, $0< \beta \leq 1$ is an exponent determining
the distribution of local relaxation processes. $\beta$ corresponds
to a stretched exponential exponent in real time. For
smaller value of $\beta$, the distribution
of local relaxation frequencies about the dominant process at
$\omega_{g} = 1/s$ is more smeared out.  If $\beta =1$, then
we will have a single overdamped oscillator of damping
time $s$- Eq.~(\ref{CD}) then corresponds to the response
function of an oscillator (e.g. Eq.~(\ref{normalres}))
with no quadratic-in-$\omega$ (inertia) term.

The mobile liquidlike component of the sample
``freezes'' at lower temperatures into a
solid glass. This, e.g., can be triggered by the quenching of dislocations
or other defects at low temperatures. In Ref.~\onlinecite{barometer},
marked superfluid-like behavior was indeed observed only
in samples which contained grain boundaries - no superfluid
behavior was noted in crystals of high quality.

This inverse susceptibility of the glass can be approximated
by a stretched set of overdamped oscillators. This
is the physical content of Davidson-Cole form of Eq.~(\ref{CD}).
The Davidson-Cole form \cite{Phase1,Phase2,Phase3} fits the
dielectric response data of glasses very well.
More sophisticated nearly universal response functions,
which may improve on the analysis that we give here,
were reported in Refs.~\onlinecite{Nagel,Dixon}.
In what follows, we will embark on an initial
Davidson-Cole analysis of the $^{4}$He data.
If we invoke the Davidson-Cole \cite{Phase1,Phase2,Phase3}
form for $g_{gl}$ (Eq.~(\ref{CD})) in
the sum of all possible intermediate
contributions of the glass to the
total response of the oscillator, then,
from Eq.~(\ref{Dyson}),
\begin{eqnarray}
\chi^{-1}(\omega) =
[\alpha - i \gamma \omega - I_{eff} \omega^{2}- \frac{g_{0}}{(1- i \omega
s)^{\beta}}].
\label{central_glass}
\end{eqnarray}

It is easy to verify, by employing
Eqns.~(\ref{periodg}, \ref{dampingg}, \ref{Qfac}), that
for very high $T$ (where $s$ is small)
Eq.~(\ref{central_glass}) predicts a period
\begin{eqnarray}
P = \frac{4 \pi I_{eff}}{\sqrt{4 [\alpha - g_{0}] I_{eff}
- (\gamma + g_{0} \beta s)^{2}}},
\label{ti}
\end{eqnarray}
and a Q-factor
\begin{eqnarray}
Q = \frac{\sqrt{(\alpha - g_{0}) I_{eff}}}{\gamma + g_{0} \beta s}.
\label{Qi}
\end{eqnarray}   Similarly, at very low $T$ (where $s$ diverges),
the corresponding quantities are given by Eqns.~(\ref{periodg},
\ref{dampingg}, \ref{Qfac}) with no change in the parameters
given therein.  A comparison of the two limits of high ($s \to 0$)
and low ($s \to \infty$) temperatures reveals that the $Q$ factor is higher at
low $T$ while the period $P$ is higher at very high temperatures $T$.

At intermediate temperatures (where $s$ is of the order of the period of
the underdamped oscillator), the angle is a superposition
of a many modes stretched about $\omega = 1/s$. From
Eqns.~(\ref{de}, \ref{ft}),
\begin{eqnarray}
\theta(t) = \frac{1}{2 \pi}
\int d\omega ~
[\alpha - i \gamma \omega - I_{eff} \omega^{2} \nonumber
\\ - \frac{g_{0}}{(1- i \omega
s)^{\beta}}]^{-1} \int dt'~\tau_{ext}(t') \exp[- i \omega (t-t')].
\label{gfinalg}
\end{eqnarray}
Eq.~(\ref{gfinalg}) represents the final form for the
angular evolution for all $T$.
Here, $\tau_{ext}(t')$ is the value of the external torque at
time $t'$. For a delta function torque, $\tau_{ext}(t') = A \delta(t')$,
the last integral simplifies. Within our general formulation,
and in Eq.~(\ref{gfinalg})
in particular, the observed dependence of the period on the
external parameters is perhaps not as surprising
as demanding a superfluid fraction strongly depend on
the oscillation velocity: For example, different initial
velocities would correspond to different values for
the internal integral 
$\int dt'~\tau_{ext}(t') \exp[- i \omega (t-t')]$.
This integral would multiply the total
susceptibility in all cases [whether
we have a glass transition, a supersolid
transition, or any other transitions].
In a treatment which is more detailed
than the one which we outline here, the transient modes
and the consequent form of the response function $g$ would also depend on
external parameters such as initial rotation velocity.

We anticipate that when the relaxation time 
is similar to the period of the
underdamped oscillator, the dissipation will be maximal.
This will lead to the largest magnitude in the imaginary
component in the singularities of $\chi^{-1}$.
Here, the glassy components respond with the same
frequency as the ``normal'' component. At both much
higher or much lower temperatures, they merely
change the net effective spring constant
but do not lead to additional modes
which closely interfere with
the oscillations of the
``normal'' (non-glass)
part. This leads
to a strong decoherence
and to a sizable energy loss.

We can examine the $T < T_{0}$ (the glass phase)
and the $T= \infty$ limits and perturbations
about them when
no dissipation is present- $\gamma =0$.
In these limits, we have ideal
harmonic oscillators with no dissipation
and consequently $Q^{-1} =0$. When we expand about
the $T=\infty$ limit and allow for a
small $s$, then as seen from Eq.~(\ref{Qi}),
$Q^{-1}$ becomes small but finite (nonzero).
Similarly, when $T=T_{0}^{+}$, where
$s$ is extremely large but finite,
$\chi$ has singularities off the real $\omega$ axis.
In Eq.~(\ref{gfinalg}), this small imaginary component
now leads to a small dissipation.
\bigskip

\begin{figure}[tb]
\begin{center}
\includegraphics[width=7.5cm,angle=0]{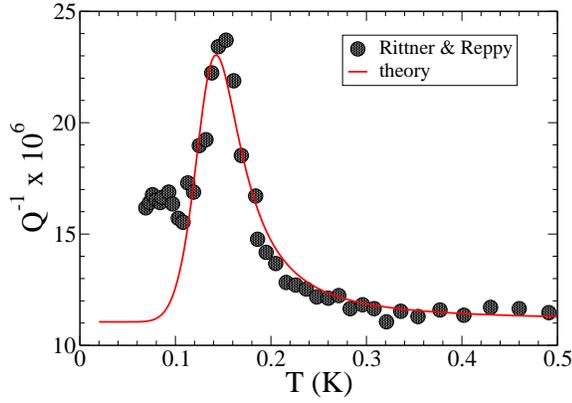}
\end{center}
\caption{(Online color) The dissipation (inverse of the quality factor)
is shown as a function of temperature
for solid \HeFour\ at 27 bar. The circles are data from
Rittner and Reppy \cite{Sophie06} and the line is our
fit above $T>0.1$ K for a simple glass model,
Eqns.~(\ref{xp}, \ref{Q-1p}), with the functional form
$Q^{-1} = A s [1+(2\pi f_0 s)^2]^{-1} + Q_\infty^{-1}$.
The fit parameters are $A=0.02776$ sec$^{-1}$,
oscillator frequency $f_0=184.2313$ Hz,
asymptotic inverse quality factor
$Q_\infty^{-1}=11.055\cdot 10^{-6}$,
glass relaxation time ${s} = 1.3537 e^{\Delta/k_B T}$ $\mu {\rm s}$,
and activation barrier $\Delta/k_B=0.9202$ K.
\newline} \label{FigQ}
\end{figure}

\bigskip

\begin{figure}[th]
\begin{center}
\includegraphics[width=7.5cm,angle=0]{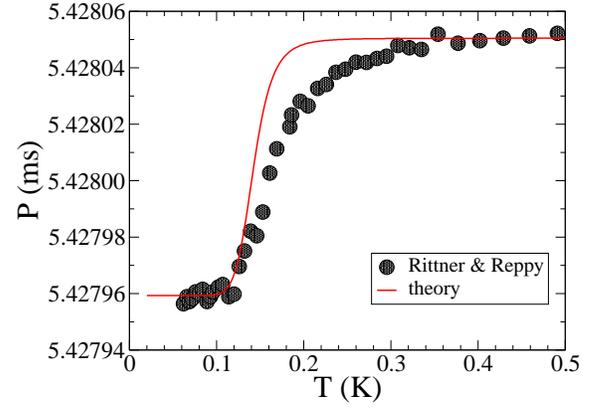}
\end{center}
\caption{(Online color) The relaxation period is shown as a function of temperature
for solid \HeFour\ at 27 bar. The circles are data from
Rittner and Reppy \cite{Sophie06} and the line  is  our
simple glass model, Eq.~(\ref{yp}), with the functional form
of Eq.~(\ref{taup}),
$P = (f_0 - {B}{f_0}^{-1} {[1+(2\pi f_0 {s})^2]^{-1}})^{-1}$.
Here the {\it only} adjustable parameter is $B=0.57$ sec$^{-2}$, otherwise
same parameters as in Fig.~\ref{FigQ}.
} \label{FigP}
\end{figure}
\bigskip


\vspace{48pt}

\begin{figure}[hb]
\begin{center}
\includegraphics[width=7.0cm]{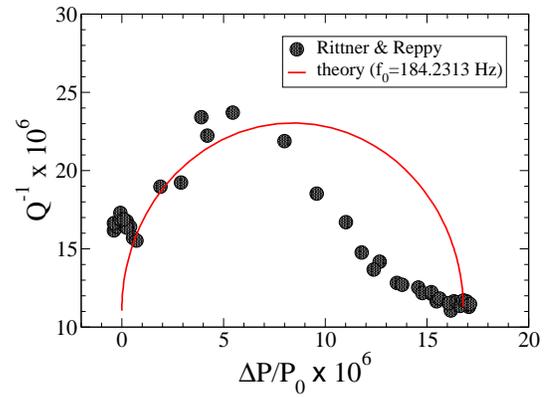}
\end{center}
\caption{(Online color) The dissipation-period plot at fixed frequency $f_0=P_0^{-1}$.
The curves of Figs.~\ref{FigQ} and \ref{FigP} are reploted against each other.
The skewness of the experimental data (circles) compared to the
semi-circle line (theory) is a well-known consequence
of a {\it real} glass exponent $\beta < 1$
[see Eqs.(\ref{CD}, \ref{central_glass})], which for simplicity
was set to $\beta=1$ in our analysis.
\newline} \label{FigQP}
\end{figure}

\bigskip

This can be made precise for $\beta =1$ where the
poles are determined by the roots of a cubic equation
in $\omega$. Here, we can simply look
for the largest magnitude of the imaginary part of all of the poles
and see when it is maximal as a function of $s$. A larger imaginary
part implies a shorter decay time and a smaller value of $Q^{-1}$.
The exact expressions are not illuminating. The asymptotic
corrections are, however, very transparent. As can be
seen from Eq.~(\ref{central_glass}), for an ideal dissipationless oscillator ($\gamma=0$), $\omega_{0}= \sqrt{\alpha/I_{eff}}$
is the pole of $\chi(\omega)$ in the low-temperature
limit where $s \to \infty$. If we expand about this pole,
$\omega= \omega_{0} + ix +y$, then we will
find that the two dominant poles attain an imaginary
component of magnitude
\begin{eqnarray}
|x| = \Big[ \frac{g_{0}}{2I_{eff}} \Big] \frac{s}{1+ (\omega_{0}s)^{2}}.
\label{xp}
\end{eqnarray}
This Lorentzian peaks, as in Fig.~\ref{FigQ},
when $s = 1/\omega_{0}$.
At this value of $s$, the poles have a maximal imaginary
component and the dissipation is the largest.
In the real dissipative oscillator (with small $\gamma >0$),
the inverse $Q$ value is given by
\begin{eqnarray}
&Q^{-1} = 2|x| \sqrt{\frac{I_{eff}}{\alpha}} + Q_{\infty}^{-1}
= \frac{1}{\sqrt{\alpha I_{eff}}} \Big[\frac{g_{0} s}{1+ (\omega_{0}s)^{2}}  + \gamma \Big] &
\nonumber \\ &&
\label{Q-1p}
\end{eqnarray}
with $Q_{\infty}^{-1} = \frac{\gamma}{\sqrt{\alpha I_{eff}}}$
the linear term in $\gamma$  (see, e.g., Eq.~(\ref{Qfac})).
By contrast, the correction to the absolute values of
the real part of the poles is
\begin{eqnarray}
y = - \Big[ \frac{g_{0}}{2 I_{eff} \omega_{0}} \Big]
\frac{1}{1+ (\omega_{0} s)^{2}}.
\label{yp}
\end{eqnarray}
The function $y$ is a monotonic function of $s$ and
thus of temperature. The period
\begin{eqnarray}
P = \frac{2 \pi}{\omega_{0} + y} 
\label{taup}
\end{eqnarray}
decreases monotonically when the temperature
is lowered, see Fig.~\ref{FigP}.

The correction for the period
due to dissipation is quadratic in $\gamma$.
This will lead to (i) a monotonic
decrease in the period as the temperature
is lowered [the increase of $y$ with lower
$T$ (higher $s$)] concurrent with (ii)
a peak in the dissipation $Q^{-1}$ at a temperature
$T^{*}$, for which $s(T^{*}) = 1/\omega_{0}$
(seen from the Lorentzian peak for $|x|$), and
not to be confused with the ideal glass transition $T_0$.
The inclusion of finite dissipation ($\gamma \neq 0$)
does not modify these trends. Generally, in glasses,
$s$ follows the Vogel-Fulcher-Tamman (VFT) form
\begin{eqnarray}
s \simeq s_{0} \exp[ DT_{0}/(T-T_{0})]   ~~~~ (T>T_{0}).
\label{VFT}
\end{eqnarray}
Here, $T_{0}$ is the temperature
at which an ideal glass transition occurs.
At temperatures $T<T_{0}$, the relaxation time
is assumed to be infinite. In some glasses,
$s$ may diverge algebraically in $(T-T_{0})$. It is
important to stress that, in principle, this may even occur
if there is no finite temperature glass transition at all and
we have a simple activated form with $T_{0}=0$ in Eq.~(\ref{VFT}),
\begin{eqnarray}
s = s_{0} \exp[{\Delta}/{k_{B}T}],
\label{act}
\end{eqnarray}
with $\Delta$ the activation barrier
which replaces the constant $(DT_{0})$ of
Eq.~(\ref{VFT}).
In Figs.~\ref{FigQ} and \ref{FigP},
we plot the results of our perturbative
calculation, Eqns.~(\ref{Q-1p}, \ref{yp}, \ref{taup}),
along with the activated form of Eq.~(\ref{act})
and compare these to the measured data
of Ref.~\onlinecite{Sophie06}.
In Fig.~\ref{FigQP} we show the dissipation-period plot of the same data
at fixed frequency $f_0$ instead of the conventional Cole plot,
\cite{Phase1,Phase2,Phase3} because
experimentally neither $\chi_1(\omega)$, nor $\chi_2(\omega)$
are available at fixed $T$.
Note that for a real glass there are many more parameters than
those that we fit with here. These parameters
are (i) $0<\beta \le 1$, which we set to 1
in our equations, and (ii) the ideal glass transition
temperature $T_{0} \ge 0$, which we set to zero
in the activated form of Eq.~(\ref{act}).
In fact, an exponent $\beta < 1$ can explain the skewness
\cite{Phase1,Phase2,Phase3} of the experimental
result in Fig.~\ref{FigQP} compared to the semi-circle result of our simplified
analysis with $\beta=1$ [see
Eqs.(\ref{xp}, \ref{Q-1p}, \ref{yp}, \ref{taup}].

The four relations of Eqns.~(\ref{periodg}, \ref{Qfac}, \ref{ti},
\ref{Qi}) along with the four measured values [the periods
at the high and low temperatures, the $Q$ factors
at high and low temperatures] allow us to determine the
four parameters $\alpha, \gamma, I_{eff},$ and $g$.
In fact, due to the scale invariance of the relations,
there are only three unknowns (e.g. $\alpha/I_{eff}, \gamma/I_{eff}$,
and $g/I_{eff}$) from which the fourth value can be predicted
and compared to its experimental value. Currently,
the available data centers on the region of
the transition and it is not certain
what the asymptotic low and
high temperature values of the $Q$ factors
($Q_{0}$ for $T=T_{0}^{+}$, $Q_{\infty}$ for $T \gg T_{0}$)
or the periods ($P_{0}$ for $T=T_{0}^{+}$, $P_{\infty}$ for
$T \gg T_{0}$) are. If we solve
for all unknown quantities in terms
of $\{P_{0}, P_{\infty}, Q_{\infty}\}$,
then we have
\begin{eqnarray}
\alpha/I_{eff} &=& 4 \pi^{2}
         \left( P_0^{-2} - P_{\infty}^{-2} [4 Q_\infty^{2} -1]^{-1} \right)
	\approx 4 \pi^2 / P_0
, \nonumber
\\ \gamma/I_{eff} &=& 4 \pi ( Q_{\infty} P_{\infty} )^{-1}
        ({ 4 - Q_{\infty}^{-2} })^{-1/2}
	\approx 2 \pi /( Q_\infty P_\infty )
, \nonumber
\\ g/I_{eff} &=& 4 \pi^{2}  (P_{0}^{-2} -P_{\infty}^{-2})
	\approx \alpha \, ( 1 - P_0^2/P_\infty^2 ).
\end{eqnarray}

The choice for the numerical value of
the effective moment of inertia $I_{eff}$ corresponds to
a different choice of units.
Neither the precise values of the
parameters $s_{0}$, $\Delta$, and $T_{0}$ of
Eq.~(\ref{VFT}), nor of $\beta$ in Eq.~(\ref{gfinalg}),
change qualitatively the
form of the fits that result from Eq.~(\ref{gfinalg}).
All that matters is that
at the temperature of the apparent transition $T^{*}$,
the dominant relaxation time
\begin{eqnarray}
\overline{s}(T^{*}) = 2 \pi s(T^{*}) \simeq P_{0},
\label{seem}
\end{eqnarray}
matches with the torsional oscillator period.
Putting all of the pieces together, Eq.~(\ref{gfinalg})
leads to a monotonic decrease of the rotation period
as the temperature is lowered.
Hand in hand with that, the dissipation becomes maximal
when $\bar{s}$ is equal to the period of the unperturbed oscillator.

A possible testable consequence of this scenario
is an experiment in which a torsional
oscillator containing solid \HeFour\ is
set in motion, and finally is abruptly stopped and then let go again. 
If the sample is first cooled below the glass transition while at rest, then
the vessel contains a solid glass and the torsional oscillator
will stop perfectly. Because all of the material inside
the oscillator will stop in unison with the container.
However, if the oscillator is stopped abruptly and let go 
at temperatures above the glass transition, then
the vessel contains a liquidlike component and the torsional
oscillator will wobble due to the motion of the
residual liquidlike component. 
Here, the liquidlike component will
still move even after the container
is stopped. This internal liquidlike motion will lead to
a momentum transfer to the outer torsional oscillator
once the container is let go. Hence,
the oscillator will wobble.

\section{Conclusions}
\label{conc}

In conclusion, we report on a
simple alternate proposal of the origin of the decrease in the
torsional oscillator period in solid helium.
\cite{Chan04,Kim05,Kim05b,Chan:05,Sophie06,Shirahama06,Kondo06,Sophie07}
We argue that changes in the oscillation period are
triggered by changes in damping and the effective stiffness
in the susceptibility and are not due to a supersolid state formation.
We point out that the torsional
oscillator mechanical properties can be controlled by a number of
effects such as damping, stiffness coefficient and moment of inertia
changes. The supersolid interpretation, if such a state exists,
would lead to changes in moment of inertia. This was the interpretation
taken in Refs.~\onlinecite{Chan04,Kim05,Kim05b, Chan:05}.
On the other hand, low-temperature
quenching of residual topological defects in the solidification into
a glass can also account for the current observation of a drop
in the torsional oscillator period as the temperature is reduced.
This effect not only accounts for the decrease in the rotation
period (Fig.~\ref{FigP}), but also predicts a peak in the dissipation,
see Fig.~\ref{FigQ}, as indeed
observed. \cite{Sophie06} Furthermore, our alternate glass scenario can
account for the observed dependence of the oscillator dynamics on
external parameters.

We have developed a  basic understanding of how a simplified
response of a glass
can account for the essential
physics results of the torsional oscillator experiments.
Specific heat data further show very strong
support for a low-temperature glass phase \cite{us} and effectively
rule out a simple supersolid transition by thermodynamic
measurements. Based on our analysis, we can
make the following specific
predictions which conform with the data available to date:

{\bf (1)} At low temperatures all dynamic
observables, e.g.,  $Q^{-1}$ and $P$,
will collapse on a
single curve as a function of temperature, which is
independent of the frequency of the
torsional oscillator.
This collapse follows from Eqs.~(\ref{Dyson}), (\ref{normalres}),
and (\ref{VFT}).
It reflects the freezing-out of liquidlike degrees of freedom, where the
response is controlled by the {\it normal} susceptibility $\chi_0$.
In general, the higher the resonant frequency $\omega_{0}$ is,
the higher the temperature $T^{*}$ at which this freeze out will occur:
$s(T_{*}) = 1/\omega_{0}$, as can be seen from
Eq.(\ref{VFT}), $T^{*} > T_{0}$. At temperatures $T \le T_{0}$,
all of the data collected from oscillators with different
resonant frequencies, in which all other parameters are
held fixed, should collapse onto a single curve.

{\bf (2)} In all known glasses the exponent $\beta < 1$
[Eq.(\ref{CD})] and the glass temperature $T_0 > 0$~K,
[Eq.~(\ref{VFT})], while in our analysis we assumed
for simplicity and to avoid the use of too many
adjustable parameters that $\beta=1$ and $T_0=0$~K. For more
realistic glass models one would need to relax these constraints.
Indeed, Fig.(\ref{FigQP}) vividly illustrates that
better fits may be obtained with values of $\beta<1$. Similarly, all
of the available data to date are consistent with values of $0< T_{0}
\alt 100$ mK. The more accurate detailed predictions
[albeit with more phenomenological glass parameters] follow from
Eqs.(\ref{central_glass}, \ref{gfinalg}, \ref{VFT}). We leave this
analysis for a separate discussion when more data are available.

{\bf (3)} The blockage of the annulus in the torsional oscillator cell
may drastically change
the strain field and dynamics inside the cell.
Such changes of the boundary
conditions can inhibit the presence of
various elastic modes relative to
those which are present in a system with no blockage.
For instance, azimuthal elastic modes which may be
present in an ``O-type'' cylindrical geometry
cannot appear in a blocked ``C-type'' cylindrical configuration.
More generally, any modification to the boundary
conditions, e.g.,
oscillator velocity or amplitude, sample surface or roughness, would shift the
freezing-out temperature and damping. For instance,
large amplitude vibrations may thwart the
quench of a liquidlike component 
into a glass. Such an avoidance  
of the glass transition will, 
consequently, lessen the low temperature 
variations in the oscillator dynamics. 
All of these effects will necessarily 
lead to different experimental observables  
without a need for invoking
a supersolid fraction which somehow
depends very sensitively on all of these
parameters.

As a matter of principle, our glass picture explains the
data even if a glass transition accompanies the transition
into a supersolid state. \cite{superglass,Wu07}
What we show is that whether or
not a tiny supersolid fraction is present- the changes
in the dynamics can be naturally ascribed
to a glass transition rather than to a Non Classical Rotational
Inertia (NCRI) effect.
Our results demonstrate that a low-temperature glass can
very naturally account for
the current data. What our analysis
suggests is whether or not a very small
supersolid fraction is present,
what triggers the change in period may (very naturally) be
a hallmark of the transition into a
glassy state. The changes in the torsional oscillator
period need not be accounted for by only a non clasical
rotational inertia effect. A glass transition (whether it is
classical or quantum is immaterial on this level
of our analysis) may further naturally account for the observed subtle
dependence of the effect on external parameters as well
as the concurrency of the peak in the
damping with the change in oscillator period.
It is worth noting that a
decrease in the rotational period
along with a peak in the dissipation
can be triggered by a host of many
other related effects - none of which
relies on a NCRI effect. As an
example, in the appendix we detail
an alternate model in which the dissipation
decreases by the
solidification of liquidlike (topological) defects
into a non-glassy low-temperature phase. Similarly,
any other effect which
leads to a decrease in the oscillator period
with decreasing temperature
and exhibits a peak in $Q^{-1}$
can account for the current data.

Finally, an experiment in which a torsional
oscillator holding \HeFour\ is first cooled when it is 
at rest, then set in motion and abruptly stopped
and let go, might allow further comparison
with the glass origin that we propose here
for the anomalous low temperature dynamics. 
For a glass transition scenario,
we expect (1) that at temperatures above the glass transition, 
the liquidlike component will
cause the torsional oscillator to
wobble. While (2) the same experiment performed sufficiently below the 
glass transition temperature, the glass and container
will stop in unison without wobbling.

\section{Acknowledgments}
This work was partially supported by the Center for Materials
Innovation (CMI) of Washington University, St. Louis and
by the US Dept.\ of Energy at Los Alamos National Laboratory
under contract No.~DE-AC52-06NA25396.
We are grateful to
A. F. Andreev, A. J. Leggett, M. Paalanen, Yu. Parshin, J. D. Reppy,
A. S. C. Rittner, and G. Volovik for useful discussions.

\appendix
\section{An alternate effect -
dissipation lowering upon freezing}
\label{alts}

Here, we show how the seminal
qualitative features of the current
available data \cite{Chan04,Kim05,Kim05b,Chan:05,Sophie06,Shirahama06,Kondo06,Sophie07}
can be explained by not only a glass
transition [or even by its more restrictive limiting
form, which is given by the simple activated dynamics of
Eqns.~(\ref{CD}, \ref{act})
with $\beta =1$ plotted in Figs.~\ref{FigQ} and \ref{FigP}), but also by other,
more general, "freezing" transitions.
Glasses adhere to specific, nearly
universal, response functions and
relaxation time forms - e.g. that of Eq.~(\ref{CD}). \cite{Phase1,Phase2,Phase3}
We now consider another "freezing" scenario of (topological) defects
in which, for illustrative purposes,
only the dissipation varies with
temperature. Upon cooling,
the liquidlike regions in the
solid "freeze" and lead to
a lower average dissipation.

In the dissipative response $g_{diss}$ of Eq.~(\ref{glassE}),
the dissipation  $\gamma(T) = \gamma_{He}(T) +
\gamma_{osc}$ varies with temperature while $I_{eff}=I_{osc}+ I_{ns}$ is
fixed.  Thus, a measured value of the period can be used to compute
$\gamma(T)$ and then predict the amplitude
damping rate $\kappa_{diss}(T)$ and compare it to
experiment. Here,
\begin{eqnarray}
\kappa_{diss}(T) = \frac{\gamma_{osc}
+ \gamma_{He}(T)}{2 (I_{osc} + I_{ns})} \nonumber
\\ = \frac{\sqrt{\alpha I_{eff} P^{2} - 4 \pi^{2} I_{eff}^{2}}}
{I_{eff} P},
\label{modelA}
\end{eqnarray}
with $I_{eff} = I_{osc} + I_{ns}$ now a constant quantity which is
independent of the temperature.
Similarly,
the Q factor (determined by $\kappa$) is
\begin{eqnarray}
Q_{diss} =
\frac{P \sqrt{\alpha}}{2 \sqrt{\alpha P^{2} - 4 \pi^{2}}}.
\label{qa}
\end{eqnarray}
If this alone is what occurs then
at all temperatures, the dissipation rate $\kappa$ and Q-factor will be related
to the period by Eqns.~(\ref{modelA}, \ref{qa}) parameterized
by the two constants $I_{eff}$ and $\alpha$. On the right hand sides
of Eqns.~(\ref{modelA}, \ref{qa}), only the measured
oscillator period $\kappa$ varies with temperature,
as indeed observed.\cite{Chan04,Kim05,Kim05b,Chan:05,Sophie06,Shirahama06,Kondo06,Sophie07}

The $Q$ factor monotonically increases as the temperature
decreases. This is evident from the final form of Eq.~(\ref{qa})
or even from the trivial Eq.~(\ref{Qfac}). Because $\gamma$ decreases
as the temperature is lowered, an ensuing increase of $Q$ follows.

We will now address the non-monotonic character of the dissipation.
This effect is not present in the treatment thus far.
The dissipation (probed by $Q^{-1}$) was seen \cite{Sophie06}
to exhibit a pronounced maximum in nearly the temperature range where
the oscillator period shows its most dramatic decrease.
Now, near transition points (whether those
of a defect freezing or of other origins),
fluctuations and consequent dissipation are very often
large. Here, we cannot think of only one damped oscillator
for either the quenching of defects nor
other transition scenario. To make contact with
the previous approach, we may divide the
system into external torsional
oscillators coupled to a multitude of
small damped oscillators within
the medium. If the medium
is homogeneous then the system
can be thought of as a single mode
oscillator. Otherwise, a more careful
analysis is required. In general, we
will need to diagonalize the $N \times N$
matrix $[\alpha- i \gamma\omega -\omega^{2} I ]$
with the elements $I_{ij}$, $\alpha_{ij}$ and $\gamma_{ij}$
being the parameters appearing in the equations of motion for the
angles $\{\theta_{i}\}_{i=1}^{N}$ of different elements of the solid.

In the context of defect
freezing, we might anticipate
some defects to already
be frozen (and lead to a lower value of
$\gamma$ in those volume elements)
and for other volume elements
to have fewer frozen defects and to
correspond to a higher value
of $\gamma$. This distribution of damping
coefficients leads to a solution for $\theta(t)$
which is a superposition of many modes
dispersed in frequency. The broader the
frequency dispersion- the larger the incoherence-
the smaller the value
of the $\langle \theta^{2} \rangle$ and that of the energy.
As a result, the $1/Q$ value which records the
damping of the energy may be much higher
near the transition.

We next implement the
stochastic character of the value of $\gamma$
for each volume element. For illustrative
purposes let us consider a distribution
given by

\begin{eqnarray}
P(\gamma) = \frac{1}{\sqrt{2 \pi \sigma_{\gamma}^{2}}}
\exp[-(\gamma - \gamma_{0})^{2}/(2 \sigma_{\gamma})].
\label{gg}
\end{eqnarray}

The width $\sigma_{\gamma}$ is generally a function of
temperature. It is finite within the transition region
and tends to zero outside it. Far from the transition,
we have a single value of $\gamma$
[i.e., $P(\gamma) = \delta( \gamma - \gamma_{0})$]
and our former analysis applies.
Within the transition region $\gamma_{0}$ can
be read off from the period by the use of Eq.~(\ref{periodg}).

The angle is now a superposition of all modes.  For a delta function
torque $\tau_{ext}(t) = A \delta (t)$,
\begin{eqnarray}
\theta(t) = Re \{ A \int_{0}^{\infty} d \gamma~ P(\gamma)
\theta_{\gamma} \exp[-i \omega_{0}^{\gamma} t - \kappa^{\gamma}t]  \},
\label{ttg}
\end{eqnarray}
$g(\gamma)$ the distribution
of $\gamma$ values, which we anticipate to be approximated by
Eq.~(\ref{gg}), and with $\omega_{0}^{\gamma}$
and $r^{\gamma}$ given by Eqns.~(\ref{periodg}, \ref{dampingg}).

To get an intuitive feeling for the damping triggered
by the dispersion, we can expand the argument
in Eq.~(\ref{ttg}) about $\gamma_{0}$ to obtain the
original solution (that assuming a single
mode defined by a unique $\gamma_{0}$)
multiplied by the approximate
damping factor
\begin{eqnarray}
C = \frac{1}{\sqrt{2 \pi \sigma_{\gamma}^{2}}}
\int_{-\infty}^{\infty} dz~
\exp[- iatz -\frac{zt}{2I_{eff}} \nonumber
\\ -
z^{2}/(2 \sigma_{\gamma})].
\label{Gauss}
\end{eqnarray}
Here, $z \equiv
(\gamma -  \gamma_{0})$ and
\begin{eqnarray}
a= \frac{\gamma}{I} \frac{1}{\sqrt{4 \alpha I - \gamma^{2}}}.
\end{eqnarray}
Let us first
examine an ideal situation
where no damping initially
exists ($\kappa^{\gamma}=0$).
Here,
Eq.~(\ref{Gauss}) is none other than the Fourier
transform of a Gaussian which is another
Gaussian given by a time dependence,
$C = \exp[- \frac{1}{2} a^{2} \sigma_{\gamma}^{2}
t^{2}]$.
This factor leads to an energy dissipation in time
that does not appear for a single mode dispersion:
the energy is damped by a factor of $C^{2}$ relative
to that of the nondispersive system. Dispersion
alone can lead to damping. Taking dissipation into
account leads to a similar Gaussian factor
of
\begin{eqnarray}
C = \exp[- \frac{1}{2}
\sigma_{\gamma}^{2} t^{2} (a^{2} + \frac{1}{4 I_{eff}^{2}})]
\end{eqnarray}
for the amplitude along with
an additional small correction to the phase
terms $\phi =
\omega_{0} t \to \omega_{0} t + \frac{\sigma^{2} a}{2I} t^{2}$.
This additional phase factor is negligible (over a period
of the oscillator) if
$(4 \alpha I - \gamma^{2})^{3/2} \gg 2 \sigma^{2} \gamma$.

The above Gaussian approximation is just a heuristic
motivation for visualizing how dispersion
lowers the average energy. The precise
integral is that of Eq.~(\ref{ttg}).
If the width $\sigma(T)$ peaks at the transition temperature then
the dissipation is amplified near the transition temperature
while, all along, the oscillator period is monotonic in temperature.
To conclude, we illustrated, once again, how the
torsional oscillator data can be accounted
for without invoking a supersolid transition.


\begin{thebibliography}{99}
\bibitem{explain_period} This form is what immediately transpires
for $\chi(\omega)$ with poles of constant $|\Re\{\omega\}|$.
When this general form is applied
to the standard oscillator, the period is indeed the
interval between consecutive local maxima of $\theta(t)$
[or, equivalently is twice the interval between
consecutive zeros of $\theta(t)$].
\bibitem{i_eff} Here and throughout, $I_{eff}$
is defined by the equations
of motion; it is determined by the coefficient of $\omega^{2}$
in $\chi^{-1}$. Generally, $I_{eff} = I_{osc} + I_{ns}$
with $I_{ns}$ the moment of inertia of the normal He
solid and $I_{osc}$ the moment of inertia of the torsional
oscillator chassis without the inserted helium. In the presence
of a supersolid component, $I_{eff} <I$, where $I$ is the
moment of inertia of the entire system.
Similar definitions apply to the dissipative coefficient
in $\chi^{-1}$. With $\gamma_{osc}$
the dissipative coefficient of the bare
torsional oscillator chassis the additional
dissipative contribution resulting from the inserted
helium is given by $\gamma_{He} = \gamma - \gamma_{osc}$.
\end{thebibliography}
\end{document}